\documentclass[prl,twocolumn,aps,superscriptaddress,preprintnumbers,letterpaper,nofootinbib]{revtex4}
\usepackage{amssymb}
\usepackage{amsmath}
\usepackage{rotating}
\usepackage{color}
\usepackage{graphicx}
\usepackage{latexsym,epsfig,bm,psfrag,subfigure}
\usepackage[bookmarksnumbered,bookmarksopen,colorlinks,citecolor=blue,linkcolor=blue]{hyperref} 
\definecolor{bobcatgreen}{rgb}{0.3,0.6,0.1}

\begin{document}
\renewcommand{\vec}{\boldsymbol}
\newcommand{\mc}{M_{\mathrm{3}}}
\newcommand{\mn}{M_{\mathrm{4}}}
\newcommand{\mnc}{M_{\mathrm{34}}} 
\newcommand{\mr}{M_{\mathrm{R}}}
\newcommand{\gma}{\gamma}
\newcommand{\gmast}{\gamma_\ast}
\newcommand{\pc}{\vec{p}_{c}}
\newcommand{\pn}{\vec{p}_{n}}
\newcommand{\bes}{{}^{7}\mathrm{Be}}
\newcommand{\besst}{{}^{7}\mathrm{Be}^\ast}
\newcommand{\be}{{}^{8}\mathrm{B}}
\newcommand{\het}{{}^{3}\mathrm{He}}
\newcommand{\hef}{{}^{4}\mathrm{He}}
\renewcommand{\S}[2]{{}^{}S_{#2}}
\renewcommand{\P}[2]{{}^{}P_{#2}}
\newcommand{\gone}{g_{(\S{2}{1/2})}}
\newcommand{\aone}{a_{(\S{2}{ 1/2})}}
\newcommand{\hone}{h_{(\P{2}{3/2})}}
\newcommand{\htwo}{h_{(\P{2}{1/2})}}
\newcommand{\V}[1]{\vec{V}_{#1}}
\newcommand{\fdu}[2]{{#1}^{\dagger #2}}
\newcommand{\fdd}[2]{{#1}^{\dagger}_{#2}}
\newcommand{\fu}[2]{{#1}^{#2}}
\newcommand{\fd}[2]{{#1}_{#2}}
\newcommand{\T}[2]{T_{#1}^{\, #2}}
\newcommand{\e}{\vec{\epsilon}}
\newcommand{\es}{\e^{*}}
\newcommand{\cw}[2]{\chi^{(#2)}_{#1}}
\newcommand{\cwc}[2]{\chi^{(#2)*}_{#1}}
\newcommand{\cwf}[1]{F_{#1}}
\newcommand{\cwg}[1]{G_{#1}}
\newcommand{\ke}{p}
\newcommand{\kest}{p_{\ast}}
\newcommand{\kc}{k_{C}}
\newcommand{\upartial}[1]{\partial^{#1}}
\newcommand{\dpartial}[1]{\partial_{#1}}
\newcommand{\etae}{\eta }
\newcommand{\etab}{\eta_{B}}
\newcommand{\etaest}{\eta_{ \ast}}
\newcommand{\etabst}{\eta_{B\ast}}
\newcommand{\vecpt}[1]{\hat{\vec{#1}}}
\newcommand{\uY}[2]{Y_{#1}^{#2}}
\newcommand{\dY}[2]{Y_{#1 #2}}
\newcommand{\llangle}{\langle\langle}
\newcommand{\rrangle}{\rangle\rangle}

\def\lsim{\mathrel{\rlap{\lower4pt\hbox{\hskip1pt$\sim$}}
    \raise1pt\hbox{$<$}}}         
\def\gsim{\mathrel {\rlap{\lower4pt\hbox{\hskip1pt$\sim$}}
    \raise1pt\hbox{$>$}}}         
\newcommand{\xilin}[1]{\textcolor{red}{\bf XZ: #1}}
\newcommand{\drp}[1]{\textcolor{bobcatgreen}{\bf DRP: #1}}
\newcommand{\kmn}[1]{\textcolor{blue}{\textbf{[KMN: #1]}}}

\title{$S$-factor and scattering-parameter extractions from $\het +\hef \rightarrow \bes + \gamma$}

\author{Xilin Zhang} \email{zhang.10038@osu.edu}
\affiliation{Department of Physics, The Ohio State University, Columbus, Ohio 43210, USA}
\affiliation{Physics Department, University of Washington, Seattle, WA 98195, USA} 
\author{Kenneth M.~Nollett} \email{nollett@mailbox.sc.edu}
\affiliation{Department of Physics, San Diego State University,
5500 Campanile Drive, San Diego, California 92182-1233, USA} 

\author{D.~R.~Phillips} \email{phillid1@ohio.edu}
\affiliation{Institute of Nuclear and Particle Physics and Department of
Physics and Astronomy, Ohio University, Athens, OH\ \ 45701, USA}
\affiliation{Institut f\"ur Kernphysik, Technische Universit\"at Darmstadt, 64289 Darmstadt, Germany}
\affiliation{ExtreMe Matter Institute EMMI, GSI Helmholtzzentrum f{\"u}r Schwerionenforschung GmbH, 64291 Darmstadt, Germany}

\date{\today\\[20pt]}

\begin{abstract}
Previous studies of the reaction $\het +\hef \rightarrow \bes + \gamma$ have mainly focused on providing the best central value and error bar for the $S$ factor at solar energies. Experimental measurements of this capture reaction at higher energies, the $\het$-$\hef$ scattering phase shifts, as well as properties of $\bes$ and its excited state, have been used to constrain the theoretical models employed for this purpose. Here we show that much more information than was previously appreciated can be extracted from angle-integrated capture data alone. We use the next-to-leading-order (NLO) amplitude in an effective field theory (EFT) for $\het +\hef \rightarrow \bes + \gamma$ to perform the extrapolation. At this order the 
EFT describes the capture process
using an s-wave scattering length and effective range, the asymptotic properties of $\bes$ and its excited state, and short-distance contributions to the $E1$ capture amplitude. We extract the multi-dimensional posterior of all these parameters via a Bayesian analysis that uses capture data below 2 MeV. We find that properties of the $\bes$ ground and excited states are well constrained. The total $S$ factor $S(0)= 0.578^{+0.015}_{-0.016}$ keV~b, while the branching ratio for excited- to ground-state capture at zero energy, $Br(0)=0.406^{+0.013}_{-0.011}$, both at 68\% degree of belief. This $S(0)$ is broadly consistent with other recent evaluations, and agrees with the previously recommended value $S(0)=0.56 \pm 0.03$ eV b, but has a smaller error bar. We also find significant constraints on $\het$-$\hef$ scattering parameters, and we obtain constraints on the angular distribution of capture gamma rays, which is important for interpreting experiments. The path forward for this reaction seems to lie with  better measurements of the scattering phase shift and $S(E)$'s angular dependence away from zero energy, together with better understanding of the asymptotic normalization coefficients of the ${}^7$Be bound states' wave functions. Data on these could further reduce the uncertainty on $S(0)$.  \end{abstract}
\maketitle

\section{Introduction}
The continuing interest in the $\het +\hef \rightarrow \bes + \gamma$ capture reaction since the 1960s~\cite{CHRISTY196189} is mostly driven by its importance to solar neutrino physics \cite{Robertson:2012ib} and primordial nucleosynthesis \cite{skm1993,nollett00,serpico04}.  The cross section can be measured directly at the \mbox{100-500 keV} energies relevant for
the Big Bang, but not at the corresponding energies for the Sun, which are around \mbox{20 keV}.  Those lower energies are not presently accessible in the laboratory due to the exponential suppression of the cross section by the Coulomb barrier.  Historically, the cross sections used in solar models have mainly been extrapolated to low energy by potential models in which data above $\approx 100$ keV were used to set a spectroscopic factor \cite{parker63,parker86,parker91,adelberger1998}; more recent efforts follow the same approach but use energy dependences based on microscopic models \cite{kajino1987,Adelberger:2010qa,Iliadis:2016vkw}.  Further inputs to the extrapolants have included $\het$-$\hef$ scattering data and $\bes$ bound-state properties.  
Concise reviews of theoretical calculations and evaluations before 2011 can be found in Refs.~\cite{Adelberger:2010qa,Nollett:2001ub}. Additional evaluations, measurements, and calculations have emerged since~\cite{neff2011,Kontos:2013qoa,Bordeanu:2013coa, deBoer:2014hha,Iliadis:2016vkw, Higa:2016igc,Dohet-Eraly:2015ooa,Premarathna:2019tup}.

We demonstrate here that the existing data on the total capture cross section tightly constrain more aspects of the reaction dynamics than a spectroscopic factor. Our analysis of those data yields quite small uncertainties on the s-wave elastic scattering parameters and the asymptotic normalization coefficients (ANCs) of the final states.  

The  framework that we use, known as Halo  effective field theory (EFT) was developed in Refs.~\cite{vanKolck:1998bw,Kaplan:1998tg,Kaplan:1998we,Bertulani:2002sz,Bedaque:2003wa,Hammer:2011ye,Rupak:2011nk,Canham:2008jd,Higa:2008dn,Ryberg:2013iga,Zhang:2017yqc} and is reviewed in Ref.~\cite{Hammer:2017tjm}. Our Halo EFT treatment of $\het(\hef,\gamma){}^7{\rm Be}$ has $\het$ ($J^\pi=\frac{1}{2}^+$) and $\hef$ ($0^+$) as fundamental degrees of freedom and $\bes$ (ground state, GS, $\frac{3}{2}^-$) and $\bes^\ast$ (excited state, ES, $\frac{1}{2}^-$) as shallow p-wave  bound states of the two. The EFT expansion is based on the observation that the lowest excitation energies of $\het$ and $\hef$ (5.5 and 19.8 MeV respectively) are much larger than the $\bes$ GS and ES binding energies (1.6 and 1.2 MeV). From the former we infer a high-momentum scale $\Lambda$ of about $200$ MeV, while from the latter we take the low-momentum scale $Q$ to be $70$--$80$ MeV. This associates the s-wave effective range, $r_0 \approx 1$ fm, with short-distance physics, but means we can examine center-of-mass energies $E$ up to about 2 MeV and still have the relative momentum in the initial $\het$-$\hef$ state fall within the purview of the EFT.

In that energy range the ratio $Q/\Lambda \approx 0.4$ is used to systematically expand both scattering and reaction amplitudes; the truncation error at a particular order can thus be estimated. We have tested that such an expansion is consistent with the physics of this process by fitting the EFT capture amplitude to previous model calculations \cite{Buck85, Buck88, Kim:1981zzb, Nollett:2001ub}. These fits give us a good understanding of the systematics of the EFT, and we then apply Bayesian analysis to the direct capture data  to extract the multi-dimensional probability distribution function (PDF) of the EFT parameters. This makes it straightforward to compute the PDF of other quantities predicted by the EFT, e.g., angular asymmetries and extrapolated $S$ factors.
   
There have also been two other recent applications of Halo EFT to this reaction~\cite{Higa:2016igc,Premarathna:2019tup}. The power counting employed in these works encodes a slightly different hierarchy of mechanisms to that used here. However, our EFT amplitude agrees with that of Ref.~\cite{Higa:2016igc} in the limit that the s-wave shape parameter is zero. The differences between our approach and that of Refs.~\cite{Higa:2016igc,Premarathna:2019tup} lie in the way that data is handled. 
Perhaps most significantly we do not include existing ${}^3$He-${}^4$He scattering phase shifts in our analysis because their errors are poorly quantified. A consistent treatment of $\het$-$\hef$ scattering is beyond the scope of this work~\cite{Adelberger:2010qa}. We also take two steps to treat correlations between data points more carefully than Refs.~\cite{Higa:2016igc,Premarathna:2019tup} did.
First, we choose either prompt or activation data from each experiment and omit the ERNA activation data so as to avoid including correlated results in our fit. Second, we include extra parameters in our analysis to account for the common-mode error in each data set. 

In the following, we first review the formulae for the EFT amplitude for $E1$ capture and discuss the power counting for this system. We then validate our power counting and amplitude: we show that it captures the low-energy behavior of several models that have been used to describe the $\het(\hef,\gamma)\bes$ reaction. The next sections discuss our Bayesian formalism and data selection, and  present our results for $S(E)$ and $Br(E)$. We then discuss other outputs of our analysis, including: constraints on EFT parameters, most notably s-wave scattering parameters; the impact of different data sets; concomitant predictions for the angular dependence of the $S$ factor, and observables that are correlated with $S(0)$. We close with a summary. 
A preliminary and abbreviated version of our results appeared in Ref.~\cite{Zhang:2018qhm}. 

\section{Formalism for $E1$ capture} 
\label{sec:E1form}
The EFT for this reaction is similar to the one constructed in our previous studies of $\bes + p \rightarrow \be + \gma$~\cite{Zhang:2017yqc, Zhang:2014zsa, Zhang:2015ajn, Zhang:2015vew}. 
The major technical differences are that here there are no core excitations and a simpler spin structure. Of course, the energy scales are also different. The NLO $S$ factor for $E1$ capture to ${}^7$Be and ${}^7$Be$^*$ can be expressed as \cite{Zhang:2017yqc}
\begin{eqnarray}
S_{_{\P{2}{3/2}}}(E) &=&  \frac{e^{2\pi \etae}}{e^{2\pi\etae}-1}  \frac{8\pi}{9}  \left(e\, Z_{eff}\right)^{2}  \kc \omega^3 C_{(\P{2}{3/2})}^{2} \times \notag \\ 
&&  \left(\mid \mathcal{S}(\gma) \mid^{2}+2 \mid \mathcal{D}(\gma) \mid^{2}\right) \ , \label{eqn:sfactormaster1} \\ 
S_{_{\P{2}{1/2}}}(E) &=&  \frac{e^{2\pi \etae}}{e^{2\pi\etae}-1}  \frac{4\pi}{9}  \left(e\, Z_{eff}\right)^{2}  \kc \omega_\ast^3 C_{(\P{2}{1/2})}^{2} \times \notag \\ 
&&  \left(\mid \mathcal{S}(\gmast) \mid^{2}+2 \mid \mathcal{D}(\gmast) \mid^{2}\right)  \ .\label{eqn:sfactormaster2}
\end{eqnarray} 
Here, $\kc\equiv \alpha_\mathrm{em} Z^2 \mr$ with $\mr$  the reduced mass of the $\het$-$\hef$ system, i.e., $\mn \mc/(\mn+\mc)$  with $\mn$ and $\mc$ as the masses of two nuclei, $\alpha_\mathrm{em} \equiv e^2/(4\pi)$ in Heaviside-Lorentz units, and both charges are $Z=2$; the well-known Sommerfeld parameter $\eta\equiv \kc/\ke$ with $\ke\equiv \sqrt{2\mr E}$; and the ``effective'' charge for the E1 transition, $Z_{eff} \equiv \left(Z/\mn-Z/\mc\right) \mr$. The energy of the photon produced in the reaction is denoted by $\omega \equiv E+ B$ and $\omega_\ast \equiv E+ B_\ast$ with GS and ES binding energies $B=1.5874$ and $B_\ast=1.1583$ MeV relative to the $\het$--$\hef$ threshold \cite{AME2016II,tunl-eval-567} (the corresponding binding momenta are $\gma \equiv \sqrt{2\mr B}$ and $\gmast \equiv \sqrt{2\mr B_\ast}$). The factors $C_{(\P{2}{3/2})}^{2}$ and $C_{(\P{2}{1/2})}^{2}$ are the squared ANCs of the $\het$-$\hef$ p-wave configurations in the GS and ES \cite{Zhang:2017yqc}. The different factors, $8/9$ and $4/9$, are due to the different multiplicity in the final states \cite{Zhang:2017yqc}. 

The other components in the two formulae are reduced matrix elements of the E1 transition between initial s- and d-wave states and final bound states, $\mathcal{S}$ and $\mathcal{D}$. Their $\gma$ or $\gmast$ dependence is made explicit to differentiate between the two reaction channels, while their dependence on other variables is left implicit.  
At NLO, the s-wave matrix elements are composed of the well-known external capture contributions together with short-distance pieces proportional to the parameter $\overline{L}$. The d-wave has only external capture. For capture to the $\bes$ GS we have
\begin{eqnarray}
&&\mathcal{S}(\gma)  \equiv \int_0^{\infty} dr W_{-\eta_{B},\frac{3}{2}}(2\gma r)  r  \Big[ C_{\etae,0}\cwg{0}(\ke,r)\frac{1}{\mathbb{N}_{\phi}(\ke)}  \notag \\ 
&& \hspace{0.4cm} +  \frac{\cwf{0}(\ke,r)}{C_{\etae,0}\,\ke} \frac{ {\rm Re}\mathbb{N}_{\phi}(\ke) }{ \mathbb{N}_{\phi}(\ke) } \Big]
 -\frac{\sqrt{3}}{2} \frac{\overline{L}}{\gma \Gamma(2+\etab)} \frac{1}{ \mathbb{N}_{\phi}(\ke)} ,  \label{eqn:Sdef}  \\ 
&& \mathcal{D} (\gma)  =   \int_0^\infty dr  {W_{-\eta_{B},\frac{3}{2}}(2\gma r)}   r  \frac{\cwf{2}(\ke,r)}{C_{\etae,0}\, \ke} \ .  \label{eqn:Ddef} 
\end{eqnarray}
Here the factor 
\begin{equation}
\mathbb{N}_{\phi}(\ke) = C_{\etae,0}^{2}\, \ke (\cot\delta_{0}-i)
\end{equation}
contains the s-wave scattering phase shift $\delta_0$ and $C_{\eta,l}\equiv 2^{l}e^{-\frac{\pi}{2}\eta}|\Gamma(l+1+i\eta)|/\Gamma(2l+2) $ with the subscript $l$ denoting orbital angular momentum. Our EFT amplitude reproduces the effective-range expansion (ERE) for $C_{\etae,0}^{2} \ke \cot \delta_0$~\cite{Higa:2008dn}:
\begin{equation}
\mathbb{N}_{\phi}(\ke)\approx -\frac{1}{a_0} + \frac{1}{2} r_0 \ke^2 - \frac{1}{4} \mathcal{P}_0 \ke^4 - 2 \kc H(\eta), 
 \end{equation}
 with the shape parameter $\mathcal{P}_0=0$ at NLO and $a_0$ and $r_0$  the scattering length and effective range.   In the coordinate-space overlaps that appear in Eqs.~(\ref{eqn:Sdef}) and (\ref{eqn:Ddef}), $W_{-\eta_{B},\frac{3}{2}}(2\gma r)$ is the asymptotic (with Coulomb) p-wave bound-state wave function with $\eta_B \equiv \kc/\gma$, while $\cwf{l}(\ke,r)$ and $\cwg{l}(\ke,r)$ are the 
standard regular and irregular Coulomb functions in the $l$th partial wave \cite{Zhang:2017yqc}.
For capture to the ES the corresponding $\mathcal{S}(\gmast)$ has $\gma$ changed to $\gmast \equiv \sqrt{2\mr B_\ast}$ and $\overline{L}$ changed to $\overline{L}_\ast$ in Eq.~(\ref{eqn:Sdef}), while $ \mathcal{D}(\gmast)$ is still found from Eq.~(\ref{eqn:Ddef}). 

We now discuss the EFT power counting that produces the amplitudes \eqref{eqn:Sdef} and \eqref{eqn:Ddef}. The leading-order (LO) p-wave scattering amplitude in Halo EFT already involves both the scattering volume and effective range terms in the effective-range expansion~\cite{Bedaque:2003wa}. In the E1 matrix element only the p-wave amplitude at the bound-state poles matters; the EFT expansion can be arranged so that this quantity receives no corrections beyond LO~\cite{Phillips:1999hh,Hammer:2011ye}. Meanwhile, the s-wave scattering amplitude is counted as $\sim 1/Q$ in the Coulomb-free case~\cite{vanKolck:1998bw,Kaplan:1998tg,Kaplan:1998we} and for situations where Coulomb is weak~\cite{Kong:1999sf}, i.e. $\eta \ll 1$, because $H(\eta) \sim Q$ there. In these contexts the term $\frac{1}{2} r_0 p^2$ in the effective-range expansion is NLO and can be treated in perturbation theory. However, for strong Coulomb ($\eta \gg 1$) $H(\eta) \sim Q^2$. If $1/a_0 \lsim Q^2$ this motivates resumming the $r_0$ piece of the effective-range-theory amplitude, yielding a LO amplitude that scales as $1/Q^2$~\cite{Higa:2008dn,Schmickler:2019ewl}. (We note that cancellations between $\frac{1}{2} r_0 \ke^2$ and ${\rm Re}~H(\eta)$ could in fact render the leading-order amplitude as large as $1/Q^3$~\cite{Higa:2016igc}.)
Our case is $\eta \sim 1$, where the best organization of the amplitude is unclear. We therefore resum the term $\frac{1}{2} r_0 \ke^2$ to  ensure that we capture its full effect. The first correction to our s-wave amplitude is then due to the $-\frac{1}{4} {\mathcal P}_0 \ke^4$ shape parameter term in the effective-range expansion. We estimate this effect to be of relative order $(Q/\Lambda)^3$ (N3LO), as per  the weak-Coulomb case~\cite{vanKolck:1998bw,Kaplan:1998tg,Kaplan:1998we}. The suppression will be less than that if the LO amplitude has a size $\sim 1/Q^2$ or $1/Q^3$~\cite{Higa:2016igc,Premarathna:2019tup}. The relative size of the short-distance effect proportional to $\overline{L}$ ($\overline{L}_\ast$ for the excited state) is independent  of the power counting adopted for the s-wave scattering amplitude, since $\mathbb{N}^{-1}_{\phi}(\ke)$ cancels in the ratio of the short-distance and external-capture contributions. The scaling of the co-ordinate space integrals for these two pieces of the amplitude evidences a suppression of the short-distance effect by a factor of $Q/\Lambda$, i.e. it enters the amplitude only at NLO ~\cite{Hammer:2011ye,Zhang:2017yqc} (cf. ``Model A" of Premarathna and Rupak~\cite{Premarathna:2019tup}). Note, however, that we do not linearize in $\overline{L}$ and $\overline{L}_\ast$ when the NLO amplitude is inserted into Eqs.~\eqref{eqn:sfactormaster1} and \eqref{eqn:sfactormaster2}. Higher-derivative short-distance operators are further suppressed by factors of $Q/\Lambda$ or $\omega/\Lambda$ according to their naive engineering dimension~\cite{Hammer:2011ye}.

Capture to both the ground and excited state involves the same initial state for s-waves ($\frac{1}{2}^+$), so the s-wave reduced matrix elements both depend on $a_0$ and $r_0$. 
Hence, up to NLO there are 6 parameters, $ C_{(\P{2}{3/2})}^2$ ($\mathrm{fm}^{-1}$), $ C_{(\P{2}{1/2})}^2$ ($\mathrm{fm}^{-1}$), $a_0$ (fm), $r_0$ (fm),  $\overline{L}$ (fm), and $\overline{L}_\ast$ (fm). In what follows we frequently work in terms of the sum and ratio of squared ANCs: $C^2_{\, T } \equiv C_{(\P{2}{3/2})}^2+ C_{(\P{2}{1/2})}^2 $ and $R_{(\P{2}{1/2})} \equiv  C_{(\P{2}{1/2})}^2/  C^2_{\, T } $.

We now validate our power counting by comparing the capture amplitude it produces to a number of models that have been used to describe the $\het(\hef,\gamma)\bes$ reaction. 

\section{EFT representation of other models} \label{sec:fitEFT}

The formulae (\ref{eqn:sfactormaster1})--(\ref{eqn:Ddef}) should approximate any model of $\het(\hef,\gamma)\bes$. As we did for $\bes(p,\gamma)\be$, we 
 now test this hypothesis by 
fitting the NLO EFT to three models: the potential models\footnote{We were unable to reproduce published results of these models, apparently because of insufficient information; we have used accurate physical constants and solution methods, and attempted to resolve ambiguities in the ways that best reproduced binding energies.} that we refer to as Buck85 \cite{Buck85}, and KimA \cite{Kim:1981zzb}, as well as a 
 semi-\textit{ab-initio} model we denote here as Nollett \cite{Nollett:2001ub}. That model used variational \textit{ab initio} models of $^3$He, $^4$He, and $^7$Be bound states but generated scattering correlations from a simplified version of the KimA potential~\cite{Kim:1981zzb}.  (We also examined the Buck88 model~\cite{Buck88}; it gave very similar results to Buck85.)
 
 The  ANCs of the potential models followed from imposing unit norm on the $^3$He-$^4$He channel, while those of the Nollett model arise from many-body dynamics in a unit-norm seven-body wave function.  The s-wave effective-range parameters, $a_0$, $r_0$, and $\mathcal{P}_0$ were fitted to the s-wave phase shifts generated by the models. 
For d-wave captures, we found that the Buck85 potential model gives results that differ significantly from the external capture formula, e.g., Eq.~(\ref{eqn:Ddef}).  We therefore included two higher-order contact terms in the EFT Lagrangian for initial d-wave channels to improve the fit.
 This adds one extra term to $\mathcal{D}(\gma)$: $\overline{L}_D \sqrt{\frac{\pi}{15}} \frac{ \ke^2}{\gma \Gamma\left(2+\eta_B\right)} \sqrt{1+\eta^2}\sqrt{4+\eta^2}$, as well as an analogous term to $\mathcal{D}(\gamma_*)$.
Both LECs, $\overline{L}_D$ and $\overline{L}_{D\ast}$, have units of $\mathrm{fm}^4$ and are expected to scale as $\Lambda^{-4}$. They should thus be N4LO effects of relative size $\approx (Q/\Lambda)^4 \approx 3\%$ at $E=2$ MeV  compared to LO. In order to fit the s-wave capture results of the KimA model we also consider the possibility of energy dependence in the short-distance contribution to s-to-p-wave $E1$ capture, via  couplings named $\overline{L}'$ and $\overline{L}_\ast'$ that modify $ \overline{L} \rightarrow \overline{L}+ \overline{L}' \ke^2$, and $\overline{L}_\ast  \rightarrow  \overline{L}_\ast+ \overline{L}_\ast' \ke^2 $. The corresponding contact operators are related to the lowest-order E1 contact operators~\cite{Zhang:2017yqc}, whose couplings are proportional to $\overline{L}$ and $\overline{L}_\ast$, but with $\vec{E}$ field replaced by $\vec{\partial}\times\vec{B}$. The new couplings have units $\mathrm{fm}^3$, and should be on the order of 0.04 fm$^3$, because the outgoing photon momentum is suppressed by $Q/M_R$ compared to the particle momenta. Numerically $Q/M_R \sim 10^{-1}$ has a similar size to $(Q/\Lambda)^2$, so this term is also considered an N4LO contribution. However, we emphasize that even including both this energy dependence and the d-to-p-wave contact operators proportional $\overline{L}_D$ and $\overline{L}_{D\ast}$ does not yield the complete N4LO calculation. 

\begin{figure}
\centering
\includegraphics[width=0.48\textwidth]{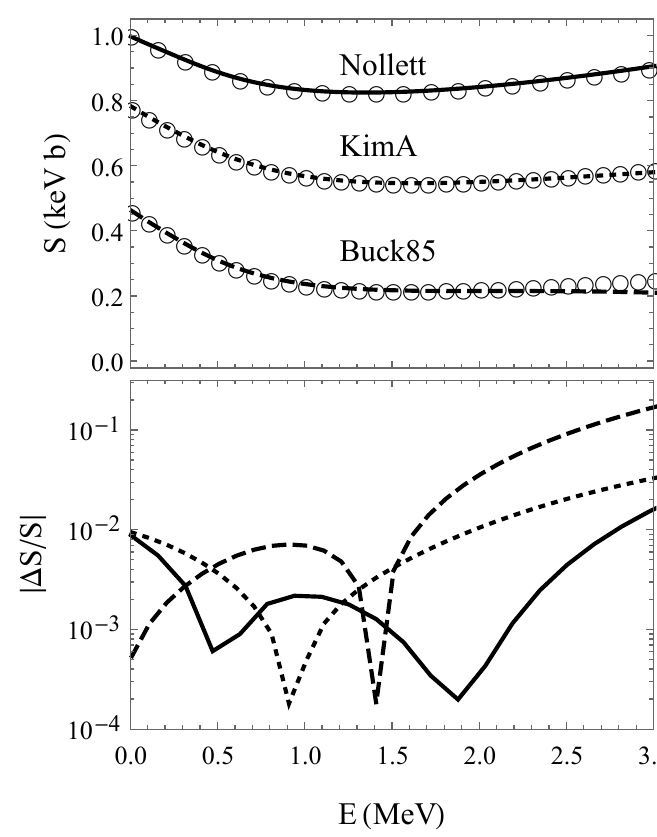}
\caption{Total $S$ factors in different potential models and EFT fits thereto.
The full $S$ factor produced by each model is represented by the open circles in the upper panel. The models, shifted for readability, are Nollett~\cite{Nollett:2001ub}, KimA~\cite{Kim:1981zzb} and Buck85~\cite{Buck85}. The solid, short-dashed, and long-dashed lines represent the partial-N4LO EFT fits to the model results, i.e., the fits whose optimal parameter values are listed in Table~\ref{tab:EFTparaModels}. The lower panel shows, on a log-linear scale, the fractional discrepancy vs energy, $\Delta S(E)/S(E)$, between each full model result and the corresponding EFT fit: Nollett (solid), KimA (short dashed), and Buck85 (long dashed).} \label{fig:FitModelTotal}
\end{figure}

The fitted parameter values are shown in Table~\ref{tab:EFTparaModels}, where it is evident that the potential models as we implemented them do not reproduce the measured binding energies. The quality of fits for the total $S$ factor can be seen in Fig.~\ref{fig:FitModelTotal}: the disagreement shows a general trend of increasing with energy, and the maximum is about $1\%$ at $E=2$ MeV. 
We find that the  ANCs are much larger than those in our previous $\bes$ capture study, indicating that the p-wave `effective range' may contain an additional fine-tuning and scale $\sim 1/\gamma$ here~\cite{Bertulani:2002sz}. Since the p-wave ANCs are comparable, $R_{(\P{2}{1/2})} \approx 0.4$, any fine-tuning would have to be present in both $J$ channels.  Meanwhile the scattering parameters are consistent with naive-dimensional-analysis estimates:  $a_0$ varies from 20 to $40$ fm, while $r_0$ is around 1 fm, corroborating the assignment of $\Lambda \approx 200$ MeV. $\mathcal{P}_0$ tends to be negative with a magnitude around 0.4 or 0.5 $\mathrm{fm}^3$. 

These numbers are consistent with capture models that treat $s$-wave ${}^3$He-${}^4$He scattering as hard-sphere scattering~\cite{Tombrello:1963zz,Williams1981,descouvemont04-bbn}. 
These match scattering data qualitatively and have $a_0=24$--$28$ fm and $r_0=1.0$ fm.
The global $R$-matrix fit of Ref.~\cite{deBoer:2014hha} corresponds to $a_0=34$ fm and $r_0=1.0$ fm; we find similar $a_0$ and $r_0$ values when we use  the AZURE2 code \cite{azure2-prc} to perform our own $R$-matrix analysis of different ${}^3$He-${}^4$He scattering data sets.

In our fits to the various potential models the NLO short-distance contributions to s-to-p-wave transitions, $\overline{L}$ and $\overline{L}_\ast$, are also consistent with the EFT estimate $\sim 1/\Lambda$. Their nearly identical size is consistent with the bound states being spin-orbit partners in the potential models. 

In contrast to the NLO scattering parameters, $\overline{L}_D$ and $\overline{L}_{D\ast}$ are considerably larger than naive-dimensional-analysis estimates in the pure potential-model cases. For Buck85 their contribution to the total cross section is $20\%$ at 2 MeV: without them, i.e., in a strict NLO calculation, the disagreement between the EFT and this model is $30\%$ at $E=2$ MeV for d-wave captures to the GS and ES.  There is a similar, although not as severe, discrepancy between the strict NLO result and the KimA model.  
Examining the Buck85 and KimA wave functions reveals that these large d-wave short-distance constants occur because in these models roughly the entire $r< 5$ fm part of the matrix element cancels out. This is because of the nodal structure imposed phenomenologically on the wave functions, which was intended to capture the main effects of nucleon-exchange antisymmetry \cite{baye85,aurdal70}.  The ultimate size of the $r < 5$ fm piece of the dipole matrix element is sensitive to the placement of the nodes and the amplitude of the oscillation between them~\cite{baye85}, and these models do not necessarily treat the physics that drives this cancellation reliably. 
In contrast, the partially \textit{ab initio} Nollett calculation implements full antisymmetry for the ${}^7$Be wave function. It yields less-complete cancellation in the one-body piece of the matrix element and hence smaller $\overline{L}_D$ and $\overline{L}_{D\ast}$ are needed in that case. 
We also note that the inferred $\overline{L}',\, \overline{L}_\ast'$ is a surprisingly large  0.5 fm$^3$ for the KimA model (the Nollett and Buck85 results are markedly smaller). This is because KimA's unphysically small $a_0=18$ fm, combined with the imposed nodal structure of its wave function, yields a noticeably different energy dependence than Eq.~(\ref{eqn:Sdef}).

Because the short-distance parameters $\overline{L}_D$, $\overline{L}_{D\ast}$, $\overline{L}' $ and $\overline{L}_\ast'$  are large compared to naturalness expectations for some models, below we study whether including them, i.e., performing a ``partial-N4LO'' calculation, significantly affects the results for $S(E)$ that we obtain from the experimental data.

\begin{table}
\begin{ruledtabular} 
   \begin{tabular}{cccc}
 & Buck85 \cite{Buck85} & KimA \cite{Kim:1981zzb} & Nollett \cite{Nollett:2001ub} \\ \hline  
 $ C^2_T $    (fm$^{-1}$)         & 30.33 & 29.22 & 21.01\\ \hline
 $ R_{(\P{2}{1/2})} $  & 0.4197 & 0.4192 & 0.4002\\ \hline
 $a_0$  (fm)              & 36.97 & 18.27 & 29.48\\ \hline 
 $r_0$  (fm)               & 0.9726 & 0.9979 & 0.9723 \\ \hline  
 $\mathcal{P}_0 $  (fm$^3$)  & -0.3688 & -0.08666 & 0.5227 \\ \hline 
 $\overline{L}$        & 0.9018 & 0.6434 & 0.9546 \\ \hline 
 $\overline{L}_\ast$   & 0.9079 & 0.6334 & 0.9772  \\ \hline 
 $\overline{L}'$        & 0.09125 & 0.5311 & 0.2240 \\ \hline  
 $\overline{L}_\ast'$   & 0.07964 & 0.5465 & 0.2366 \\ \hline
 $\overline{L}_D$ (fm$^4$)     & -4.541 & -1.950 & 0.5124 \\ \hline 
 $\overline{L}_{D\ast}$ (fm$^4$) & -4.844 & -3.096 & 0.3444 \\ \hline 
$B$ (MeV)             & 1.608  & 1.656 & 1.587 \\ \hline 
$B_\ast$ (MeV)      & 1.163 & 1.192 & 1.158 \\ \hline 
\end{tabular}  \caption{EFT parameters obtained from partial-N4LO fits to models from the literature. Quantities are total ANC squared ($C^2$), the ratio of ES ANC squared to $C_T^2$, the scattering length, effective range, and shape parameters as well as s-to-p-wave short-distance parameters, $\overline{L}$ and $\overline{L}_\ast$ and d-to-p-wave short-distance parameters $\overline{L}_D$ and $\overline{L}_{D\ast}$. The last two rows are the binding energies of $\bes$'s GS and ES. } \label{tab:EFTparaModels}
\end{ruledtabular}
\end{table}

\section{Bayesian analysis and data selection}
We now take the results (\ref{eqn:sfactormaster1})--(\ref{eqn:Ddef}), together with data from a number of recent experiments~\cite{Brown:2007sj,NaraSingh:2004vj,Bemmerer:2006xa, Confortola:2007nq,Gyurky:2007qq,Costantini:2008ub,DiLeva:2009zz,Kontos:2013qoa,Bordeanu:2013coa}
and employ Bayesian analysis~\cite{SiviaBayesian96, Schindler:2008fh,Furnstahl:2014xsa}---implemented via Markov-Chain-Monte-Carlo (MCMC) sampling\footnote{We find it useful to implement parallel tempering~\cite{Earal2005PCCP} to handle local maxima in the likelihood as a function of $\mathbf{g}$.}---to obtain probability distribution functions (PDFs) for the EFT parameters. This follows the path of our work on $\bes(p,\gamma)^8\mathrm{B}$ and details can be found in Refs.~\cite{Zhang:2015ajn,Zhang:2015vew}. The first goal is to obtain the posterior PDF of the parameter vector $\vec{g}$ given data, $D$, our theory, $T$, and prior information, $I$.

To account for the common-mode errors (CMEs) in the data we introduce data-normalization corrections, $\xi_J$ (with $J=1$ to $N_\mathrm{exp}$, the number of $S$-factor experiments, which ultimately equals six) in addition to the six (eleven) fitted EFT parameters at NLO (partial-N4LO) that constitute the vector $\vec{g}$. The vectors $\vec{g}$ and $\vec{\xi}$ contain the full set of parameters that we want to constrain. For the EFT parameters we take flat prior distributions with a range that comfortably encompasses natural values (see below). If we then 
use gaussian priors for the $\xi_J$'s, we can (omitting obvious step functions) write the desired PDF as $ {\rm pr} \left(\vec{g},\{\xi_J\} \vert D;T; I \right) \equiv c\, \exp\left(-\chi^2/2\right) $, where $c$ is chosen so that the PDF is properly normalized and 
\begin{eqnarray}
\chi^2 & \equiv &   \sum_J^{N_\mathrm{exp}} \bigg\{\sum_{j=1}^{N_{s,J}} \frac{\left[ (1-\xi_J)S(\vec{g}; E_{Jj})-D_{Jj}\right]^2}{ \sigma_{Jj}^{2}} + \frac{\xi_J^2}{\sigma_{c,J}^2} \bigg\} \notag \\ 
 && + \sum_{l=1}^{N_{br}} \frac{\left[ Br(\vec{g}; E_l)-\tilde{D}_l\right]^2}{ \sigma_{br, l}^{2}}.    \label{eqn:PDFdef}
\end{eqnarray}
Here $J$ indexes the different $S$-factor experiments and $j$ labels the $j$th data point in a particular experiment, $D_{Jj}$, taken at energy $E_{Jj}$, and with point-to-point uncertainty $\sigma_{Jj}$. $S(\vec{g};E_{Jj})$ is then the EFT prediction for the total $S$ factor at that energy.  Meanwhile experiment $J$'s quoted CME is $\sigma_{c,J}$.
The second part of the first term then comes from the priors for the $\xi_J$ normalization parameters.  Note that although we use the notation $\chi^2$ in Eq.~(\ref{eqn:PDFdef}) to emphasize the similarity to the usual goodness of fit parameter, the quantity defined there does not follow a $\chi^2$ distribution because of the CMEs.

There are six total $S$-factor data sets, here labeled Seattle (S) \cite{Brown:2007sj}, Weizman \cite{NaraSingh:2004vj}, Luna (L) \cite{Bemmerer:2006xa, Confortola:2007nq,Gyurky:2007qq,Costantini:2008ub}, Erna  \cite{DiLeva:2009zz}, Notre Dame \cite{Kontos:2013qoa}, Atomki \cite{Bordeanu:2013coa}.  Their stated CMEs,  $\sigma_{c,J}$, are, respectively, 3\%, 2.2\%, 2.9\%, 5\%, 8\%, and 5.9\%. The lowest energy data are from Luna, while the points above 1.5 MeV are mostly from Erna with three from Atomki.  
Wherever possible, we have used activation data from these experiments following the same reasoning as in Ref.~\cite{Adelberger:2010qa}: the prompt and activation data are correlated whenever both exist in the same experiment, but the prompts have an additional source of systematic uncertainty due to the photon emission anisotropy assumed in their analysis.  We do however use recoil data from Erna.  We also use the prompt measurements from Notre Dame. 

In order to ensure that we are fitting data within the domain of validity of the EFT we only use data below 2 MeV, which yields a total of 59 $S$-factor and 32 $Br$ data points. Fig.~\ref{fig:SBrvsENLON4Lv2FullData} displays the data sets. The bump in the $S$-factor data at $E=3$ MeV is due to $E2$ capture to the GS through a $\frac{7}{2}^-$ f-wave resonance; it does not affect the partial waves that matter at astrophysical energies, and  we estimate its contribution in the region of our analysis, $E < 2$ MeV, to be  $<0.1$\%.
We discuss our data selection and provide a full data listing in the supplemental material, where we correct some misprints in the published literature (cf. the EXFOR database \cite{exfor} entry for Ref.~\cite{Confortola:2007nq}). 

The second sum in Eq.~(\ref{eqn:PDFdef}) brings branching-ratio data into our analysis.  The total number of $Br$ measurements is $N_{br}$. The $l$th data point is $\tilde{D}_l$, with point-to-point uncertainty $\sigma_{br, l}$. The EFT result is then $Br(\vec{g}; E_l)$, the ratio of excited-state to ground-state cross sections. There are four modern data sets on this branching ratio: from Seattle, Luna, Erna, and Notre Dame. Since the $S$-factor data are mostly either activation or recoil and the biggest CMEs in $S$---target thickness, detector geometries, etc.---largely cancel for branching ratios, we assume no correlation between $S$-factor and $Br$ data\footnote{The exception is ND, where we include both total $S$-factor and $Br$ values from the same prompt-gamma counting but both statistical and systematic errors are large compared with other data.  
%
}.  And, since CMEs largely cancel in $Br$, we do  not include an analogue of the $\xi_J$'s for these data, so they do not need to be grouped by experiment.

\section{Results for $S(E)$ and $Br(E)$}

We define the EFT parameter space via the quadrature-sum and ratio-of-squared ANCs, $C_T^2$ and $R_{\P{2}{1/2}}$, together with the sum and difference of the short-distance LECs: $\overline{L}_{T,\delta} \equiv \overline{L} \pm \overline{L}_\ast$. We take flat priors for these parameters, with ranges considerably larger than those suggested by naive dimensional analysis: $ 0 < C_T^2 \leq 100$ $\mathrm{fm}^{-1}$, $ 0 < R_{(\P{2}{1/2})} < 1 $, $0 \leq a_0 \leq 75$ fm, $0< r_0 \leq 10$ fm, $-20 \leq \overline{L}_T \leq 20$ fm, $-20 \leq \overline{L}_\delta \leq 20$ fm. We have taken $a_0$ to
be positive since this is indicated by microscopic models~\cite{walliser83}, phenomenological treatments of scattering data~\cite{Tombrello:1963zz,Williams1981,deBoer:2014hha}, and our own AZURE2 fits.  Without such a prior, our analyses produce alternate maxima of the posterior corresponding to negative $a_0$. 
We also require that there is no resonance below 1.6 MeV and no bound state shallower than 1.6 MeV. Note that choosing flat priors on 
$C_{(\P{2}{3/2})}^2$ and $C_{(\P{2}{1/2})}^2$ instead of $C_T^2$ and $R_{P_{1/2}}$ results in a different weight in the PDF, but the two choices give nearly identical results, because the Jacobian is approximately constant for the ANCs that are ultimately permitted by the data.

There are no further corrections to  our Halo EFT at N2LO, so the dominant truncation errors should be N3LO and N4LO terms. At N3LO the shape parameter, $\mathcal{P}_0$ is incorporated into the $\mathbb{N}_{\phi}$ factor in Eq.~(\ref{eqn:Sdef}) for the GS and the analogous expression for the ES. 
In addition to our NLO analysis we do a partial-N4LO analysis that includes selected terms associated with short-distance physics at N4LO (see above). This allows us to assess the impact of truncating the EFT at NLO. In the partial-N4LO calculation we include the $\overline{L}'$ and $\overline{L}_\ast'$ couplings describing the energy dependence of the s-wave-capture's contact terms and  also 
 the contact terms in the d-wave channel parameterized by $\overline{L}_D$ and $\overline{L}_{D\ast}$. We take flat priors for these five N3LO and N4LO parameters. The priors are chosen to cover ranges that are much larger than the naive dimensional analysis and that cover the fit results for models and existing literature on the s-wave ERE~\cite{Blatt:1949zz,Kamouni:2007hhh}. They are: $-0.6 \leq  \mathcal{P}_0 \leq 0.6 \;\mathrm{fm}^3$; 
 $-1\leq \overline{L}',\, \overline{L}_\ast'\leq 1 \; \mathrm{fm}^3$;
 $ -10 \leq \overline{L}_D, \overline{L}_{D\ast} \leq 10 \; \mathrm{fm}^4$.
%

Turning now to the empirical data, sampling the total $\chi^2$ defined by Eq.~(\ref{eqn:PDFdef}) reveals no improvement going from NLO to partial-N4LO.
We also evaluate the Bayes factor~\cite{Sivia1996,Furnstahl:2014xsa} using the Savage-Dickey ratio~\cite{Dickey:1971} and find a Bayes factor for NLO over partial-N4LO of about 6. This constitutes ``substantial" evidence against the partial-N4LO calculation for these observables in this energy range~\cite{JeffreysBook}. We therefore quote NLO as our main result.
Full results for the partial-N4LO calculation are given in the Supplemental Material.

\begin{figure}
\centering
\includegraphics[width=0.45\textwidth]{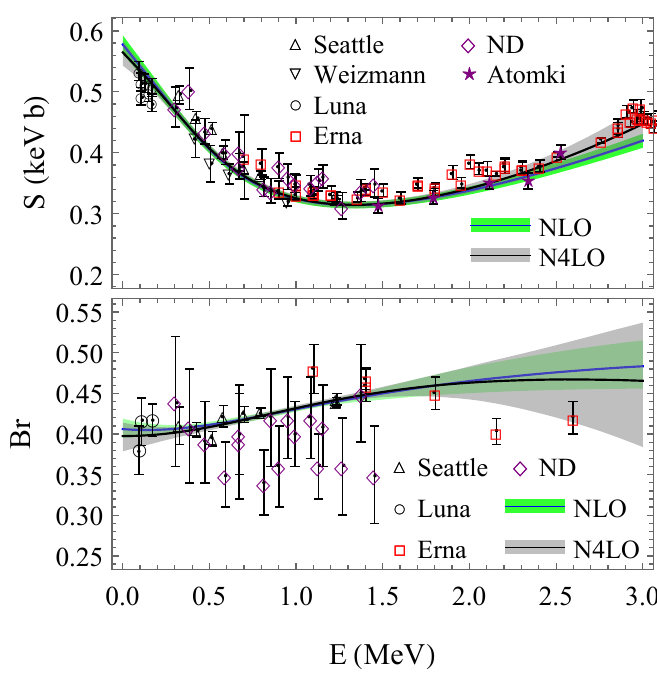}
\caption{Total $S$-factor and branching-ratio results. The data sets are denoted as in the legend and are from Luna~\cite{Bemmerer:2006xa, Confortola:2007nq,Gyurky:2007qq,Costantini:2008ub}, Erna~\cite{DiLeva:2009zz}, Seattle~\cite{Brown:2007sj}, Weizman~\cite{NaraSingh:2004vj}, Atomki~\cite{Bordeanu:2013coa}, and Notre Dame (ND)~\cite{Kontos:2013qoa}. The green (gray) band shows the 68\% interval for $S(E)$ and $Br(E)$ in our NLO (partial-N4LO) analysis. The mean is denoted by the blue (black) line.} \label{fig:SBrvsENLON4Lv2FullData}
\end{figure}

Fig.~\ref{fig:SBrvsENLON4Lv2FullData} shows the 68\% interval for $S(E)$ and $Br(E)$ in our NLO (green) and partial-N4LO (gray) analyses. The NLO (partial-N4LO) mean is denoted by the blue (black) line. Note that the data is shown without any re-scaling by the factors $\xi_J$, so Fig.~\ref{fig:SBrvsENLON4Lv2FullData} under-reports how well our final result reproduces the data. If we adopt values for the $\xi_J$'s that maximize their posterior PDF then the distribution of $\chi^2$'s of our MCMC sample peaks at 82, or 1.04 per degree of freedom.

\begin{figure}
\centering
\includegraphics[width=0.4\textwidth]{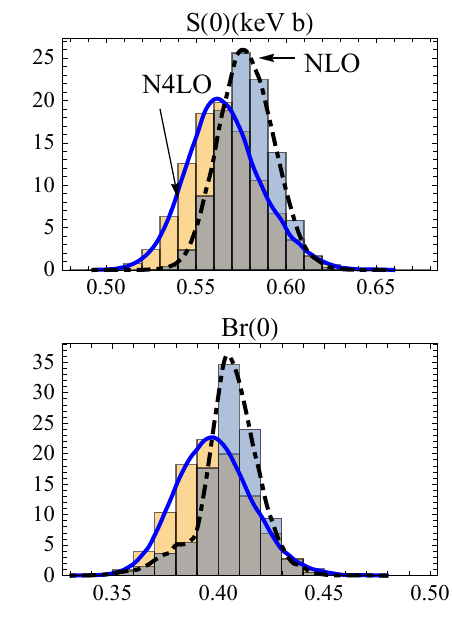}
\caption{One-dimensional distributions obtained for NLO (grey histograms, black dash-dotted lines) and partial-N4LO (yellow histograms, blue solid line) analyses. The histograms are generated from our MCMC samples and the lines are reconstructed smooth distributions. The upper panel shows $S(0)$ and the lower panel shows $Br(0)$.} \label{fig:s0br0disNLON4LOv2FullData}
\end{figure}

The one-dimensional PDFs for $S(0)$ and $Br(0)$ are shown in Fig.~\ref{fig:s0br0disNLON4LOv2FullData}. For $S(0)$ the NLO EFT result is  $S(0)= 0.578^{+ 0.015}_{-0.016}$ keV~b while the partial-N4LO result is $S(0)= 0.565^{+ 0.019}_{-0.022}$ keV~b. The 68\% interval for $S(0)$ at NLO thus also encompasses the impact of higher-order EFT corrections. We also compute $S$ at the $20$ keV representative of the solar environment, and get $S(20~\mathrm{keV})=0.570 \pm 0.015$ keV~b. This is compatible with  a recent inference from measurements of the solar-neutrino flux: $S(20~\mathrm{keV})=0.548 \pm 0.054$ keV~b~\cite{Takacs:2015yua}.

The recommended $S(0)$  from fitting mainly the same data in Ref.~\cite{Adelberger:2010qa} is $0.56\pm 0.02\mathrm{(exp)} \pm 0.02\mathrm{(theory)}$ keV~b, which is consistent with our result, but has an uncertainty that is almost a factor of two larger. The more recent R-matrix analysis of deBoer et al., which included $\het$-$\hef$ scattering data, gave $0.542 \pm 0.011 \mathrm{(MC\, fit)} \pm 0.006 \mathrm{(theory)} ^{+0.019}_{-0.011}\mathrm{(phase\, shift)}$ keV~b~\cite{deBoer:2014hha}, while Iliadis et al. quote $0.572\pm 0.012\mathrm{(exp)} \pm 0.013\mathrm{(theory)}$ keV~b~\cite{Iliadis:2016vkw}. deBoer et al.'s value is lower than ours by 1.4  standard deviations (adding all errors in quadrature); it may be relevant that the scattering data included in their fit greatly outnumber the $S$-factor data. Our result is consistent with that of Iliadis et al., but has a slightly smaller error bar. More recently, Premarathna and Rupak considered two different Halo EFT power countings of the ${}^3$He-${}^4$He scattering and capture reactions~\cite{Premarathna:2019tup}. 
Neither power counting is exactly the same as ours, but in practice the NLO amplitude we used to obtain our main results agrees with the leading-order 
``Model A" amplitude of Ref.~\cite{Premarathna:2019tup}. Our analysis therefore matches most closely the ``Model A* II" calculation of Ref.~\cite{Premarathna:2019tup} which found $S(0)=0.551^{+0.021}_{-0.014}$ keV~b.  That 68\% interval overlaps ours, with the different central values potentially explained by some differences in the way we treated data (see above) and Premarathna and Rupak's inclusion of the shape parameter ${\mathcal P}_0$ as a free parameter in their ``Model A" fit. The earlier Halo EFT analysis by Higa et al.~\cite{Higa:2016igc} included scattering data and so is harder to compare to ours; it yielded $S(0)=0.558 \pm 0.008$ keV~b, which was updated to $S(0)=0.550^{+0.009}_{-0.010}$ keV~b in Ref.~\cite{Premarathna:2019tup}.

The NLO and partial-N4LO numbers for $Br(0)$ are 
$Br(0)=0.406^{+0.013}_{-0.011}$ and $Br(0)=0.397^{+0.017}_{-0.019}$ respectively. In this case adding the higher-order terms increases the uncertainty by about 50\%.

\section{The impact of different data sets}

\begin{figure}
\centering
\includegraphics[width=0.4\textwidth]{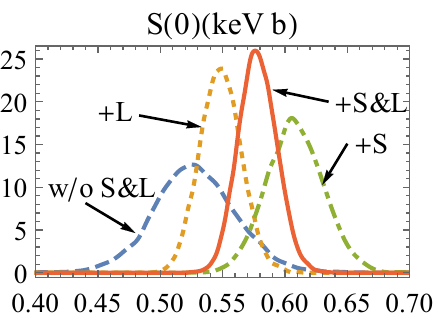}
\includegraphics[width=0.4\textwidth]{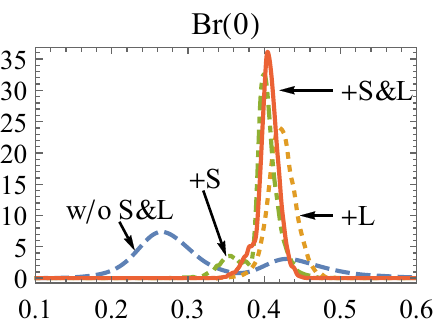}
\caption{The impact on $S(0)$ and $Br(0)$ of analyzing subsets of our full data set. The blue-dashed ``w/o S \& L" curve is the result if Seattle and Luna data are both excluded, while the orange-dashed ``+ L" and green-dot-dashed ``+ S" curves show the result when that analysis is supplemented with the Luna and Seattle data  respectively. The solid curve ``+S \& L" curve is our final NLO result, including all data sets described above.  } \label{fig:dataimpact1dimdis2}
\end{figure}

We can use our NLO EFT analysis to study how choosing particular data sets affects the results for $S(0)$ and $Br(0)$. We did three new extractions: the first excludes the precise/low-energy data from Seattle (S) and Luna (L) on $S(E)$ and $Br(E)$, while the other two include either the Seattle or the Luna data. These are to be compared to the above results, which employ all data. 
Fig.~\ref{fig:SBrvsENLON4Lv2FullData} suggests that the $Br$ measurements from ND and Erna provide only weak constraints on the extrapolation to $E=0$ and 
Fig.~\ref{fig:dataimpact1dimdis2} confirms that without either the Seattle or Luna data sets the PDF for $S(0)$ is quote broad, while that for $Br(0)$ has two peaks. 
Fig.~\ref{fig:SBrvsENLON4Lv2FullData} makes it clear that the Luna data set then adds constraints at very low energy, while the Seattle experiment provides more precise measurements at higher energies. These two data sets, either singly or more powerfully in combination, remove the $Br(0) \approx 0.25$ solution---a solution that would be quite surprising in light of existing quantum mechanical models. Individually, they also increase the $S(0)$ median value from $0.52$ to $0.55$ (Luna) and $0.60$ (Seattle). The one-dimensional PDFs for $S(0)$ in the upper panel of Fig.~\ref{fig:dataimpact1dimdis2}
then show a consistency region for $S(0)$ when all of the data sets are considered, with a narrower peak than can be obtained if the Seattle or Luna data are excluded. 


\section{EFT and normalization parameters}

The one-dimensional distributions of the EFT parameters and the $\xi_J$ from our MCMC analysis can be found in the Supplementary Material (see Fig.~\ref{fig:1dimDisNLOFullDataLargerWindow}). We also used the VEGAS \cite{Lepage:1977sw} integration algorithm to marginalize the 12-dimensional PDF down to one dimension and found good agreement with MCMC.  The NLO analysis leads to strong constraints on the ANCs and scattering parameters: there is a lot of information gained from direct-capture data as compared to the large prior windows we started with. We find (68\% intervals): 
$C_T^2 =27\pm 3~\mathrm{fm}^{-1}$, $R_{(\P{2}{1/2})} = 0.48 \pm 0.04$, $a_0 = 52 \pm 5$ fm, $r_0=0.97 \pm 0.03$ fm, $\overline{L}_T = 2.35 \pm 0.1$ fm, $\overline{L}_\delta=0.08\pm 0.08$ fm. The data supports $r_0$ and both  short-distance parameters of around 1 fm, validating the EFT power counting. (This is also close to the values obtained in the fits to potential-model output---see Table~\ref{tab:EFTparaModels}.) Note also that $\overline{L} \approx \overline{L}_*$, i.e., the short-distance physics makes approximately the same contribution to capture to both ground and excited states, as expected for spin-orbit partner states.

Including the N4LO EFT operators relaxes these constraints to varying degrees, but the central values generally remain within the 68\% interval of the NLO fit, see the Supplementary Material.  There are no strong constraints on the five N4LO parameters, although positive values are preferred for all of them. The partial-N4LO fit implies $\overline{L}_D$ and  $\overline{L}_{D\ast}$ are between $-2$ and $2$ fm. This is much more consistent with naive dimensional analysis than were the potential-model fits in Table~\ref{tab:EFTparaModels}. We note that this outcome is {\it not} a result of our choice of prior, which encompassed a markedly broader range. 

The $\xi_J$ distributions of both Seattle and Luna data sets have central values slightly outside the 68\% interval associated with the quoted CME $\sigma_{J,c}$, but they are within the 95\% interval.  Refs.~\cite{Adelberger:2010qa,Iliadis:2016vkw} also found some ground for regarding these sets as moderate outliers.  The other sets have CME distributions within or barely outside the 68\% region. 

\section{Angular asymmetry}

From our analysis of mainly activation and recoil data, we can also infer the angular asymmetry of the prompt gamma rays.   All prompt-gamma measurements of this reaction involve detectors covering limited solid angle, and published cross sections generally include a correction for angular asymmetry based on models \cite{Tombrello:1963zz,Kim:1981zzb} that amounts to 2--3\% after integration over detector solid angle \cite{Confortola:2007nq,Costantini:2008ub,Brown:2007sj,Kontos:2013qoa,nagatani69,hilgemeier88,osborne84,alexander84}.  This correction is sometimes assigned a 100\% error, which causes it to dominate the error budget of at least one experiment \cite{Costantini:2008ub}.  A desire to place this correction on a firmer footing has been expressed in the recent experimental literature \cite{Brown:2007sj,Kontos:2013qoa}.

At NLO the differential $S$ factor, ${d S}/{d \Omega}$, in the CM frame is proportional to the reaction amplitude squared, i.e., 
$$
\left\vert \mathcal{S}\right\vert^2 +2 \left\vert \mathcal{D}\right\vert^2   + \left[  2\, \mathrm{Re}\,\left( e^{i(\sigma_{2}-\sigma_{0})} \mathcal{S}^{\ast} \mathcal{D}\right) - \left\vert \mathcal{D}\right\vert^2 \right] P_2 (\cos\theta), \label{eqn:aniformula}
$$
with $e^{2i\sigma_{l}} \equiv  \Gamma(l+1+i\eta)/\Gamma(l+1-i\eta)$ the phase shift due to the Coulomb interaction in partial wave $l$.
In what follows we define the anisotropy of the cross section, $A_2$ such that  ${d S}/{d \Omega}\propto 1 + A_2 \cos^2 \theta$~\cite{Tombrello:1963zz}.
Other modulations proportional to $P_1(\cos\theta)$ and $P_3(\cos\theta)$ should---because of parity---mainly arise due to interference between E1 and M1 or E2, whereas  $A_2$ should be dominated by E1 transitions below 2 MeV. Near the $\frac{7}{2}^-$ resonance the E2 multipole could contribute, but our NLO result for $A_2$ should be robust for the energies considered here. 

\begin{figure}
\centering
\includegraphics[width=0.4\textwidth]{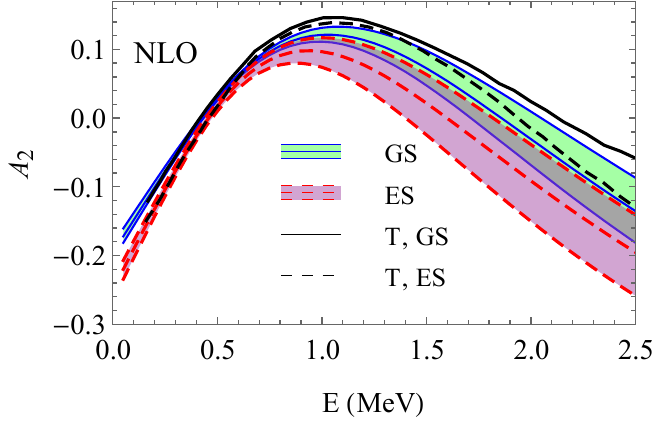}
\caption{Blue solid (red dashed) lines are the mean and boundaries of the 68\% intervals for the anisotropy $A_2$ for capture to GS (ES) at NLO in Halo EFT. The green and purple regions show the full 68\% interval for the GS and ES respectively. The black solid (black dashed) line is the result of Tombrello~\cite{Tombrello:1963zz} for the asymmetry in radiative capture to the ground (excited) state, commonly used to analyze experimental data.} \label{fig:aniNLO}
\end{figure}

We compute $A_2$ for both the $P_{3/2}$ ground state and the $P_{1/2}$ excited state, using our NLO samples, and obtain the result shown in  Fig.~\ref{fig:aniNLO}. The smooth blue (GS) and red (ES) curves are the mean values from the $A_2$ distributions at different energies, while the corresponding bands are the 68\% intervals.  The  anisotropy used to interpret most experiments~\cite{Tombrello:1963zz} (see Fig.~4 therein) is also shown: it lies at the upper edge, or just outside, our 68\% band at $E> 500$ keV.  
Fig.~\ref{fig:aniNLO} shows our uncertainty for $A_2$ is generically around $20\%$  below 1 MeV and grows towards higher energies.  Our results suggest that---at least away from the zero of $A_2$ around 400-500 keV---the uncertainty of the $A_2$ correction has sometimes been overestimated.  Ref.~\cite{Brown:2007sj} found that at 700--1200 keV setting $A_2=0$ produced better agreement between prompt and activation data in the same experiment than the Ref.~\cite{Tombrello:1963zz} anisotropies; our results do not support such a large deviation from Ref.~\cite{Tombrello:1963zz}. 
The only direct measurement of the anisotropy is in Ref.~\cite{kraewinkel82}, where errors are too large to provide strong constraints. 

\section{$S(0)$ and its correlants}

\begin{figure}
\centering
\includegraphics[width=0.3 \textwidth]{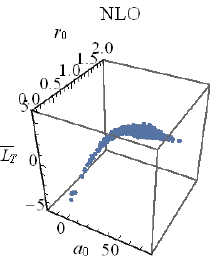}
\caption{$a_0$--$r_0$--$\overline{L}_T$ 3-dim scatter plot based on MCMC sampling.  All three axes have units of fm. } \label{fig:a_r_Lt_Corr_NLOFullData}
\end{figure}

Part of the power of Bayesian methods lies in their ability to reveal correlations between different parameters. For example, Fig.~\ref{fig:a_r_Lt_Corr_NLOFullData} shows a three-dimensional scatter plot of the NLO MCMC samples in the $a_0$--$r_0$--$\overline{L}_T$ sub-space. When projected onto the  $a_0$--$r_0$ and  $a_0$--$\overline{L}_T$ planes this structure produces hyperbolic correlations. This can be understood from  inspection of  Eq.~(\ref{eqn:Sdef}) and expressions immediately following it. At NLO in Halo EFT the data constrains the combinations of parameters $a_0 (r_0 + \mathrm{constant})$ and $a_0 ( \overline{L}_T + \mathrm{constant})$; the constant comes from the $H(\eta)$ function in the ERE and the $G_0(\ke,r)$ wave function contribution in the reduced matrix element. 

\begin{figure}
\centering
\includegraphics[width=0.45\textwidth]{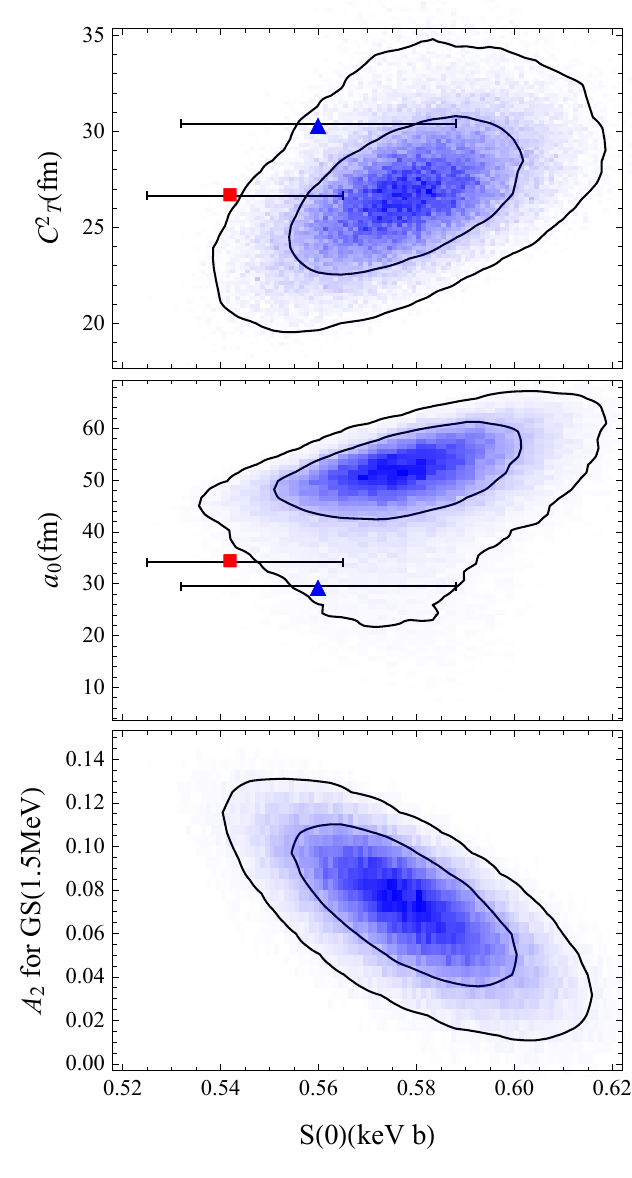}
\caption{NLO correlation of $S(0)$ with $C_T^2$, $a_0$, and $A_2$ of the ground-state transition at 1.5 MeV. In the top panel, the red square denote $C_T^2$ and $S(0)$ from Refs.~\cite{deBoer:2014hha} and~\cite{Adelberger:2010qa} (only uncertainty for $S(0)$ quoted). The values of $a_0$ and $S(0)$ from Refs.~\cite{deBoer:2014hha,Adelberger:2010qa} are also shown in the middle panel (again, only uncertainty for $S(0)$ is shown). }  \label{fig:S0CorrOthersNLO}
\end{figure}

Interesting two-dimensional correlations between $S(0)$ and other observables are shown in Fig.~\ref{fig:S0CorrOthersNLO}: with $C^2_T$, $a_0$, and the anisotropy $A_2$ of the ground-state transition at 1.5 MeV. (We chose the particular energy 1.5 MeV as representative of energies well above zero, where larger cross sections may mean $A_2$ can be measured with reasonable accuracy.)
The opacity of the histogram grows with the size of the pertinent PDF; the contours then correspond to the 68\% and 95\% regions around the mode. Interestingly,  the correlation between the ANCs and $S(0)$ is not nearly as strong as in the $\bes(p,\gamma)$ reaction. Other physics affects $S(0)$ more markedly here than in that case~\cite{Baye:2000ig}. We also show the evaluation of $S(0)$ and the ANCs from deBoer's $R$-matrix analysis~\cite{deBoer:2014hha}  (red square) and the ``Solar-Fusion II" evaluation~\cite{Adelberger:2010qa} (blue triangle): both are consistent with this correlation at the combined 68\% level, even without any consideration of ANC uncertainty.
The correlation between $S(0)$ and $a_0$ is clear in the NLO analysis; we also show where the deBoer {\it et al.} and Adelberger {\it et al.} results lie in that plane.  

Lastly, a measurement of $A_2$ near $1.5~\mathrm{MeV}$ may provide additional information on $S(0)$. 
 $A_2(1.5~{\rm MeV})$ near 0.1 clearly favors an $S(0)$ at the low end of our range, while $A_2 (1.5\mathrm{MeV})$ near zero would imply an $S(0)$ at the upper end of it---and a concomitantly larger $C_T^2$. (Similar correlation also exists between $S(0)$ and the $A_2 (1.5\mathrm{MeV})$ for the total $S$ factor.) This output from an $A_2$ measurement is independent of our discussion in the previous section of the role of $A_2$ in interpreting the prompt measurements.

\section{Summary}

We employed the Halo EFT expressions for $\het(\hef,\gamma)\bes$ at next-to-leading order to describe with reasonable accuracy the behavior obtained for the $S$ factor of this reaction in three different models. However, in order to accurately reproduce some models' $S$-factor behavior up to 2 MeV the NLO calculation must be supplemented by selected higher-order terms, namely those that encode the short-distance part of the d-to-p-wave $E1$ transition. We then considered 59 $S$-factor and 32 $Br$ data points, obtained below center-of-mass energies of 2 MeV in experiments at six different laboratories. We employed a Bayesian analysis, with broad priors on EFT parameters predicated on them being natural, to determine that (68\% degree of belief) $S(0)= 0.578^{+0.015}_{-0.016}~\mathrm{keV~b}$, given these data and the NLO Halo EFT expression.  The central value is consistent with, but the error bar approximately a factor of two smaller than, the recommendation of Ref.~\cite{Adelberger:2010qa}. Our $S(0)$ is broadly consistent with other, more recent, analyses of this reaction~\cite{deBoer:2014hha, Iliadis:2016vkw}. Our results can be used to analyze whether parameters and observables are correlated with $S(0)$, and so determine how to obtain additional or complementary constraints. We also showed the impact of different data sets on our determination of the $S$ factor at solar energies, and found that the angular asymmetry assumed in previous prompt-gamma experiments is not grossly wrong.  Further measurements of that asymmetry would address an important aspect of prompt-gamma measurements, and could also tighten constraints on the EFT parameters, thereby reducing the $S(0)$ uncertainty. 

Natural EFT parameters can be found such that the NLO expression provides a good fit to the data---after common-mode errors are accounted for---with a $\chi^2$ per degree of freedom of 1.04. The $\chi^2$ is not improved by adding the short-distance mechanisms embodied in simple potential models of $S(E)$, and so the Bayesian evidence ratio disfavors their inclusion. Perhaps most surprisingly, the NLO fit provides a strong constraint on the s-wave scattering length and effective range, $a_0 = 52 \pm 5$ fm, $r_0=0.97 \pm 0.03$ fm---without the use of any $\het$-$\hef$ scattering data. This is in marked contrast to our $\bes(p,\gamma)$ analysis where information on $a_0$ had to be taken from other experiments. The incorporation of $\het$-$\hef$ scattering information into the analysis is an important topic for further work on the use of Halo EFT and Bayesian methods in $\het(\hef,\gamma)\bes$. 

{\em Acknowledgments---}
We acknowledge James deBoer for useful discussions, and for significant help in use of the AZURE2 $R$-matrix code. We thank Gautam Rupak for useful discussions and for a careful reading of the manuscript. 
We also thank Carl Brune for valuable conversations and for encouraging us to examine this reaction in Halo EFT. This research was supported by the US Department of Energy, Office of Science, Office of Nuclear Physics under Awards DE-FG02-93ER-40756 (DRP),  DE-SC0019257 (KMN), DE-FG02-97ER-41014 (XZ), through MSU subcontract RC107839-OSU for the NUCLEI SciDAC collaboration (XZ), by the National Nuclear Security Agency under Award DE-NA0003883 (DP),
by the US NSF via grants PHY-1614460 (XZ) and PHY-1630782, N3AS FRHTP (DP), by the  ExtreMe Matter Institute EMMI at the GSI Helmholtzzentrum f\"ur Schwerionenphysik (DP), and by the US Institute for Nuclear Theory (XZ).

\bibliographystyle{apsrev}
\bibliography{He3He4CaptureForPLB-XZ-0916.bbl}

\begin{widetext}
\newpage
\section{Supplemental Material} \label{sec:SM}

%
%
%
%
%


\subsection{More details on the data sets}

The Weizman data have been multiplied by a factor of $10.52/10.44$ following Ref.~\cite{deBoer:2014hha}. We note that the statistical error bars of three Luna data points have been over-represented by a factor of two in previous analyses~\cite{deBoer:2014hha, Iliadis:2016vkw} due to mistakes in some published tables~\cite{Costantini:2008ub,exfor} (see EXFOR documentation for Ref.~\cite{Costantini:2008ub} data at \url{https://www.nndc.bnl.gov/exfor/servlet/X4sGetSubent?reqx=53982&subID=241709002}).

For the ERNA data we use the recoil data, since for that case the asymmetry appears not to be relevant and there are only a few activation data; we then exclude the ERNA activation data because of their correlations with the recoil data and with the LUNA data for which the same counting apparatus was used. For the Notre Dame data set there is no activation data and the larger errors mean that the anisotropy correction of ``less than 2\%" is roughly a quarter the size of both the systematic and the statistical errors, so we use the prompt data there.  

Finally, we note that the Seattle $Br$ data were analyzed assuming $A_2=0$. Correcting for the anisotropy of the emission of the gamma ray from capture to the ground state could affect these results since the correction should be applied to the GS capture gamma-ray but not to the isotropic cascade gamma from ES capture. Determining whether a statistically significant correction due to this effect should be applied to the Seattle $Br$ data is beyond the scope of this publication. 

The following tables list the 71 data used in our analysis, together with the point-to-point errors.

\begin{table}[h!]
\begin{ruledtabular} 
   \begin{tabular}{ccccc}
  E(MeV) & $S$(eV b) & $\delta S$ (eV b) &  $BR$ & $\delta BR$ \\ \hline  
0.3274 & 0.495 & 0.015 & 0.41 & 0.023 \\
 0.426 & 0.458 & 0.01 & 0.405 & 0.009 \\
 0.518 & 0.44 & 0.01 & 0.394 & 0.009 \\
 0.5815 & 0.4 & 0.011 & 0.422 & 0.013 \\
 0.7024 & 0.375 & 0.01 & 0.424 & 0.01 \\
 0.7968 & 0.363 & 0.007 & 0.427 & 0.005 \\
 1.2337 & 0.33 & 0.006 & 0.439 & 0.006 \\
 1.2347 & 0.324 & 0.006 & 0.443 & 0.007 \\
\end{tabular}  \caption{ Seattle data } \label{tab:seadata}
\end{ruledtabular}
\end{table}

\begin{table}[h!]
\begin{ruledtabular} 
   \begin{tabular}{ccc}
  E(MeV) & $S$(eV b) & $\delta S$ (eV b)  \\ \hline  
0.42 & 0.423 & 0.0308 \\
 0.506 & 0.382 & 0.0301 \\
 0.615 & 0.365 & 0.0162 \\
 0.95 & 0.318 & 0.00576 \\
\end{tabular}  \caption{Weizmann data } \label{tab:weidata}
\end{ruledtabular}
\end{table}

\begin{table}[h!]
\begin{ruledtabular} 
   \begin{tabular}{ccccc}
  E(MeV) & $S$(eV b) & $\delta S$ (eV b) &  $BR$ & $\delta BR$ \\ \hline  
 0.3034 & 0.475 & 0.033 & 0.44 & 0.08 \\
 0.3849 & 0.505 & 0.034 & 0.41 & 0.07 \\
 0.4742 & 0.435 & 0.021 & 0.39 & 0.05 \\
 0.5931 & 0.401 & 0.02 & 0.35 & 0.04 \\
 0.6717 & 0.378 & 0.021 & 0.4 & 0.05 \\
 0.6718 & 0.402 & 0.032 & 0.39 & 0.07 \\
 0.815 & 0.344 & 0.016 & 0.34 & 0.04 \\
 0.856 & 0.338 & 0.02 & 0.42 & 0.06 \\
 0.9028 & 0.379 & 0.021 & 0.36 & 0.05 \\
 0.9516 & 0.361 & 0.021 & 0.42 & 0.05 \\
 0.994 & 0.35 & 0.015 & 0.4 & 0.04 \\
 1.084 & 0.346 & 0.017 & 0.42 & 0.05 \\
 1.129 & 0.355 & 0.018 & 0.36 & 0.04 \\
 1.1545 & 0.361 & 0.019 & 0.41 & 0.05 \\
 1.2674 & 0.313 & 0.022 & 0.36 & 0.06 \\
 1.3741 & 0.338 & 0.018 & 0.45 & 0.06 \\
 1.452 & 0.35 & 0.023 & 0.35 & 0.06 \\
\end{tabular}  \caption{Notre Dame data } \label{tab:nddata}
\end{ruledtabular}
\end{table}

\begin{table}[h!]
\begin{ruledtabular} 
   \begin{tabular}{ccc}
  E(MeV) & $S$(eV b) & $\delta S$ (eV b)  \\ \hline  
 1.473 & 0.313 & 0.012 \\
 1.791 & 0.327 & 0.011 \\
 2.115 & 0.351 & 0.01 \\
 2.338 & 0.354 & 0.013 \\
 2.527 & 0.401 & 0.012 \\
\end{tabular}  \caption{Atomki data } \label{tab:atomdata}
\end{ruledtabular}
\end{table}

\begin{table}[h!]
\begin{ruledtabular} 
   \begin{tabular}{ccc}
  E(MeV) & $S$(eV b) & $\delta S$ (eV b)  \\ \hline  
 0.0929 & 0.534 & 0.0156 \\
 0.1057 & 0.493 & 0.0144 \\
 0.1265 & 0.514 & 0.0151 \\
 0.1477 & 0.499 & 0.0146 \\
 0.1689 & 0.482 & 0.0141 \\
 0.1695 & 0.507 & 0.0149 \\
 0.1056 & 0.516 & 0.0151 \\
\end{tabular}  \caption{Luna $S$ data } \label{tab:lunaSdata}
\end{ruledtabular}
\end{table}

\begin{table}[h!]
\begin{ruledtabular} 
   \begin{tabular}{ccc}
  E(MeV) & $Br$ & $\delta Br$  \\ \hline  
 0.171878 & 0.417 & 0.02 \\
 0.107424 & 0.415 & 0.029 \\
 0.094533 & 0.38 & 0.03 \\
\end{tabular}  \caption{Luna $Br$ data } \label{tab:lunaBrdata}
\end{ruledtabular}
\end{table}

\begin{table}[h!]
\begin{ruledtabular} 
   \begin{tabular}{ccc}
  E(MeV) & $S$(eV b) & $\delta S$ (eV b)  \\ \hline  
 0.701 & 0.393335 & 0.0690062 \\
 0.802 & 0.384988 & 0.0210953 \\
 0.902 & 0.338772 & 0.0149145 \\
 1.002 & 0.350661 & 0.0125236 \\
 1.002 & 0.33277 & 0.0107345 \\
 1.102 & 0.334043 & 0.00309299 \\
 1.102 & 0.338683 & 0.00618598 \\
 1.103 & 0.333598 & 0.0092666 \\
 1.203 & 0.333206 & 0.00682798 \\
 1.203 & 0.333206 & 0.0122904 \\
 1.353 & 0.327112 & 0.00820711 \\
 1.403 & 0.343345 & 0.00448817 \\
 1.403 & 0.339979 & 0.00897635 \\
 1.504 & 0.338643 & 0.010356 \\
 1.604 & 0.325511 & 0.00965908 \\
 1.704 & 0.348755 & 0.0108986 \\
 1.704 & 0.350572 & 0.00817395 \\
 1.804 & 0.344794 & 0.00343935 \\
 1.804 & 0.339635 & 0.010318 \\
 1.904 & 0.367645 & 0.0114633 \\
 1.955 & 0.3505 & 0.00880252 \\
 2.005 & 0.385407 & 0.0109669 \\
 2.055 & 0.373849 & 0.00921189 \\
 2.105 & 0.373511 & 0.0120487 \\
 2.156 & 0.36588 & 0.00369576 \\
 2.205 & 0.380778 & 0.0116268 \\
 2.205 & 0.377871 & 0.0116268 \\
 2.305 & 0.374302 & 0.00985006 \\
 2.306 & 0.374891 & 0.0112538 \\
 2.406 & 0.378399 & 0.00956243 \\
 2.507 & 0.396918 & 0.00398913 \\
 2.762 & 0.420305 & 0.00439124 \\
 2.857 & 0.443373 & 0.0123159 \\
 2.857 & 0.437215 & 0.0123159 \\
 2.908 & 0.467309 & 0.0103711 \\
 2.928 & 0.453491 & 0.01155 \\
 2.947 & 0.476221 & 0.0102999 \\
 2.968 & 0.458807 & 0.0120739 \\
 2.987 & 0.456732 & 0.0054158 \\
 2.988 & 0.475307 & 0.0120331 \\
 3.008 & 0.457537 & 0.0101941 \\
 3.028 & 0.454249 & 0.00956315 \\
 3.048 & 0.45279 & 0.0119155 \\
 3.068 & 0.443044 & 0.00950229 \\
 3.089 & 0.457582 & 0.0112472 \\
 3.11 & 0.434283 & 0.00826081 \\
 3.13 & 0.429449 & 0.0176486 \\
\end{tabular}  \caption{Erna $S$ data } \label{tab:ernaSdata}
\end{ruledtabular}
\end{table}

\begin{table}[h!]
\begin{ruledtabular} 
   \begin{tabular}{ccc}
  E(MeV) & $Br$ & $\delta Br$  \\ \hline  
 1.102 & 0.48 & 0.03 \\
 1.403 & 0.46 & 0.02 \\
 1.403 & 0.468 & 0.013 \\
 1.804 & 0.45 & 0.02 \\
 2.156 & 0.403 & 0.016 \\
 2.597 & 0.42 & 0.02 \\
\end{tabular}  \caption{Erna $Br$ data } \label{tab:ernaBrdata}
\end{ruledtabular}
\end{table}

\subsection{Further N4LO results}

In Fig.~\ref{fig:1dimDisNLOFullDataLargerWindow} we show the one-dimensional distributions for the EFT parameters and the experimental normalization parameters, $\xi_i$, obtained in our NLO and partial-N4LO analyses. 

\begin{figure}[h!]
\centering
\includegraphics[width=0.85\textwidth]{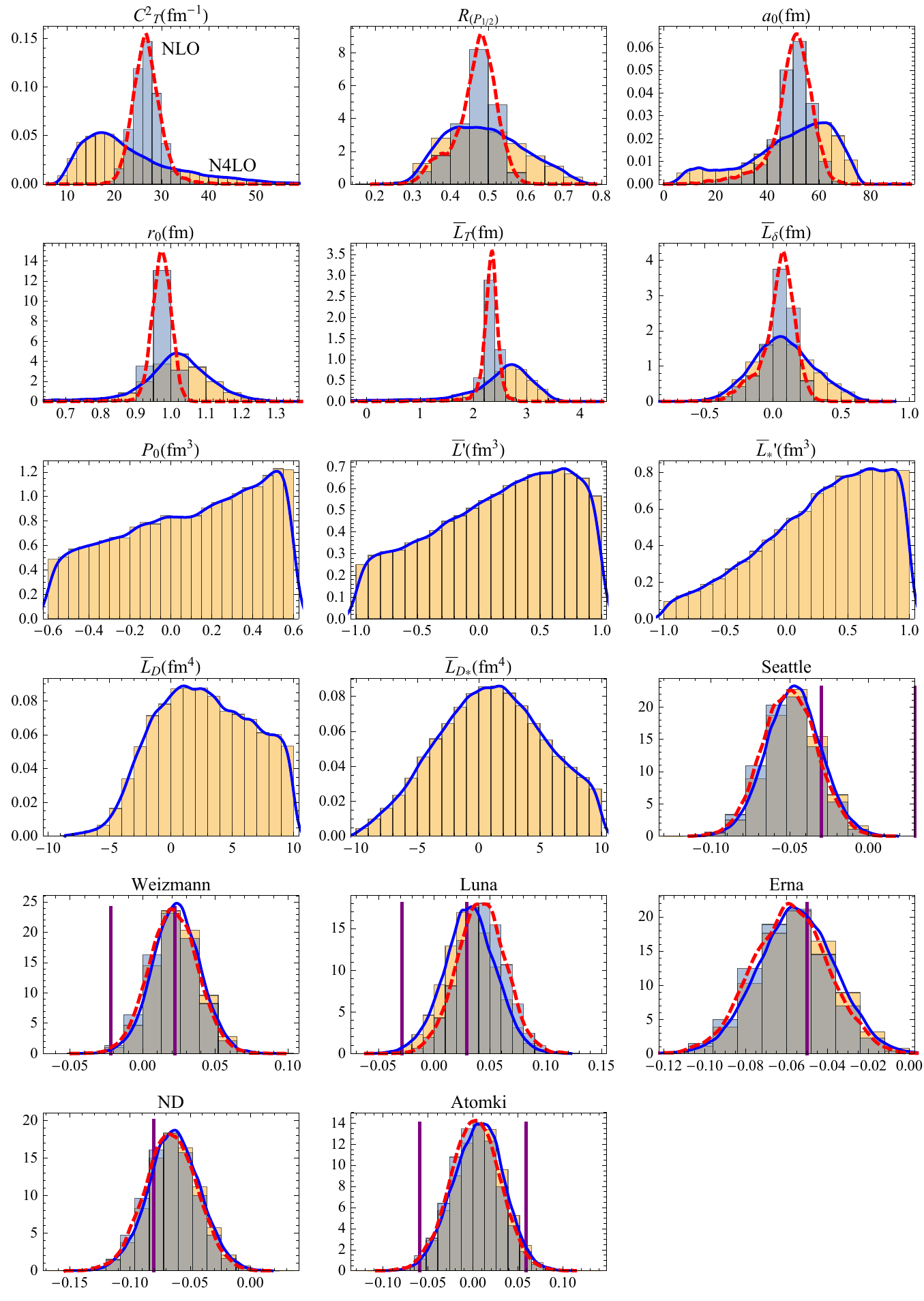}
\caption{One-dimensional PDFs for NLO (grey histograms, red dash-dotted lines) and partial-N4LO (yellow histograms, blue solid lines) analyses. The histograms come from the MCMC samples; the lines are the reconstructed smooth distributions. In the plots of the $\xi_i$ distributions the vertical purple lines show the 1$\sigma$ normalization uncertainty for each data set quoted in the corresponding publication and stated in the main text.} \label{fig:1dimDisNLOFullDataLargerWindow}
\end{figure}
\end{widetext}

The experimental data we analyzed prefer positive values of $\mathcal{P}_0$, $\overline{L}'$, $\overline{L}_\ast'$, $\overline{L}_{D}$, and $\overline{L}_{D\ast}$.
 This is in contrast to the results for potential models: Table~\ref{tab:EFTparaModels} shows  negative values for $\mathcal{P}_0$ in all models and  a sizable and negative $\overline{L}_{D\ast}$ in the two that lack explicit nucleon antisymmetry. Despite the fact that there are no strong constraints on these five additional parameters their presence in the fit does mean that $C_T^2$ ($a_0$) develops a long tail at large (small) values in the partial-N4LO PDFs.  $a_0$ also develops a second mode corresponding to ``small" $a_0$ ($< 20$ fm). 
 
\begin{figure}
\centering
\includegraphics[width=0.4\textwidth]{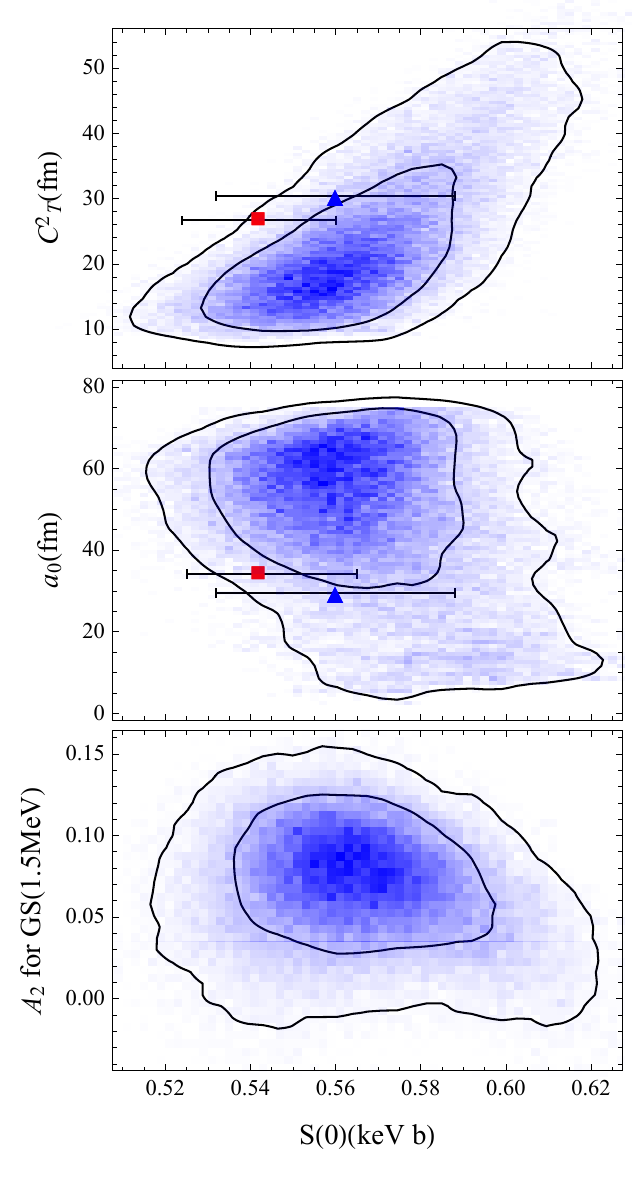}
\caption{$S(0)$ correlated with other observables. (Note $A_2$ is for the ground-state transition.) In the top (middle) panel the red square represents the $C_T^2$ ($a_0$) and $S(0)$ from Ref.~\cite{deBoer:2014hha} and the blue triangle that from Ref.~\cite{Adelberger:2010qa}. Both references, however, only provide an uncertainty for $S(0)$ and do not give an error for $C_T^2$ or $a_0$.} \label{fig:S0CorrOthersN4LO}
\end{figure}

This means that the correlation between $S(0)$ and $a_0$ is less clear in the partial-N4LO calculation, see Fig.~\ref{fig:S0CorrOthersN4LO}. If other data, e.g., from $\het$-$\hef$ scattering, eliminates the second mode then the partial-N4LO correlation between $a_0$ and $S(0)$ will remain very similar at partial-N4LO to the NLO one shown in Fig.~\ref{fig:S0CorrOthersNLO}.

\begin{figure}
\centering
\includegraphics[width=0.3 \textwidth]{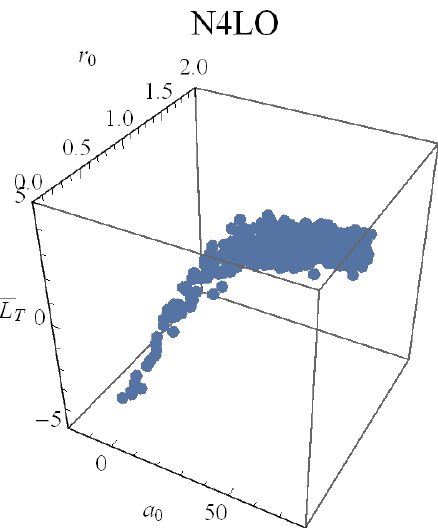}
\caption{$a_0$--$r_0$--$\overline{L}_T$ (all in units of fm) three-dimensional scatter plot based on the partial-N4LO MCMC ensemble.}
\label{fig:a_r_Lt_Corr_N4LOFullData}
\end{figure}

The inclusion of the additional N4LO parameters also broadens the three-dimensional correlation of $a_0$, $r_0$, and $\bar{L}_T$, see Fig.~\ref{fig:a_r_Lt_Corr_N4LOFullData}, which is the analog of Fig.~\ref{fig:a_r_Lt_Corr_NLOFullData} in the main text. 

\begin{figure}
\centering
\includegraphics[width=0.4\textwidth]{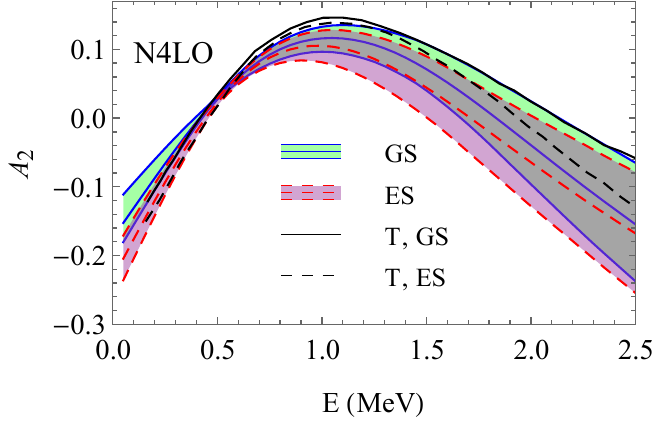}
\caption{Solid blue (dashed red) lines indicate the mean and 68\% interval for the anisotropy factor for the capture to the GS (ES) in the partial-N4LO MCMC samples. Green (purple) fills in the  68\% region. The black solid (GS) and dashed (ES) are the anisostropy calculation of Ref.~\cite{Tombrello:1963zz}. } \label{fig:aniN4LO}
\end{figure}

The anisotropy prediction at partial-N4LO is also entirely consistent with the NLO one, see Fig.~\ref{fig:aniN4LO}, which is to be compared to Fig.~\ref{fig:aniNLO}. Since the data  do not support the additional effects in the partial-N4LO calculation,  the N4LO results in this Supplemental Material have 68\% intervals that overlap the NLO 68\% intervals. The partial-N4LO intervals do tend to be broader, but that is presumably largely because the five extra parameters in that calculation are so poorly constrained by the data. 

\end{document}